\documentclass[showpacs,twocolumn,floatfix,prl]{revtex4}
\usepackage{graphicx}
\usepackage{amsfonts}
\usepackage{amsmath}
\usepackage{amssymb}
\usepackage{graphicx}%

\newcommand{\beq}{\begin{equation}}
\newcommand{\eeq}{\end{equation}}
\newcommand{\beqn}{\begin{eqnarray}}
\newcommand{\eeqn}{\end{eqnarray}}
\newcommand{\bearr}{\begin{array}}
\newcommand{\enarr}{\end{array}}

\begin{document}

\title{dc-biased stationary  transport  in the absence of dissipation}

\author{S. Denisov$^1$, S. Flach$^1$ and P. H\"anggi$^{1,2}$}
\affiliation{$^1$Max-Planck Institut f$\ddot{u}$r Physik
Komplexer Systeme, N$\ddot{o}$thnitzer Str. 38, D-01187 Dresden, Germany, \\
$^2$Universit\"at Augsburg, Institut f\"ur Physik,
Universit\"atsstrasse 1, D-86135 Augsburg, Germany}

\date{\today}
\vskip 2.cm
\begin{abstract}

We obtain stationary transport in a Hamiltonian system with ac
driving in the presence of a dc bias. A particle in a periodic
potential under the influence of a time-periodic field possesses a
mixed phase space with regular and chaotic components. An
additional external dc bias allows to separate effectively these
structures. We show the existence of a stationary current which
originates from the persisting invariant  manifolds (regular
islands, periodic orbits, and cantori). The transient dynamics of the
accelerated chaotic domain separates fast chaotic motion from
ballistic type trajectories which stick to the vicinity of the
invariant submanifold. Experimental studies with cold atoms in
laser-induced optical lattices are ideal candidates for the
observation of these unexpected findings.
\end{abstract}

\maketitle

Bridging microscopics and macroscopics remains the ultimate goal
in statistical mechanics, be it in or out of equilibrium. A
considerable amount of work has been devoted to investigate this
problem for low-dimensional stationary non-equilibrium states
\cite{Dorfman, Hoover, Brownian}. Stationary transport states
under an external bias are often connected to nonzero dissipation (phase
space volume contraction and violation of time reversal
invariance), which, in turn, is equivalent to positive entropy
production (for a recent review see Ref. \cite{Chaos}). Attempts
to keep time reversal invariance by using Nose-Hoover thermostats
\cite{Hoover} did not leave the grounds of dissipative systems and
phase space volume contraction.

The interface between microscopic dynamics and statistical
evolution and the role of nonlinearity and dissipation in
nonequilibrium stationary states is an active area of
research.  Modern experimental studies provide an
ideal testing ground to explore these problems using manipulations
with  cold atom ensembles in optical potentials \cite{cold}. The
driven pendulum, a paradigmatic system for the study of dynamical
chaos \cite{Ham}, is realized using a periodically modulated optical standing
wave, with the possibility to control the strength
of dissipation down to arbitrarily small values
\cite{cold}. For example, recent cold
atoms experiments \cite{Renzoni} study the crossover
from dissipative to Hamiltonian ratchets,
confirming theoretical predictions \cite{Yevtushenko}.

In this Letter we explore the
route of obtaining stationary transport using a dc bias in the
absence of dissipation and corresponding phase space
volume contraction.
A necessary condition for such a stationary, bias driven current to
occur is a \textit{mixed} phase space \cite{Ham} with coexisting
regular invariant manifolds and chaotic regions for the unbiased
case. A consequence of this coexistence is that
directed transport may arise locally on regular invariant
components of phase space. While the chaotic phase space regions
transform into accelerated evolution for any arbitrary small value
of the dc bias, the regular invariant submanifolds (unstable periodic
orbits, regular islands and cantori  with nonzero mean
velocities), initially embedded in the chaotic region, persist
under a finite dc bias. Moreover  \textit{unstable} periodic orbits
at zero dc bias lead to the appearance  of \textit{stable}
islands for some finite nonzero value of  the dc
bias. The presence of cantori \cite{Cantor} and the corresponding
sticking of chaotic trajectories in the vicinity of regular
manifolds is manifested by the chaotic phase space part which,
though accelerated, shows large time separation from chaotic
trajectories which are initially located far from the regular
structures.

We consider the canonical Hamiltonian model of a particle moving in a
one-dimensional space-periodic potential $U(X)=\frac{1}{2\pi}\cos(2\pi x)$ under the
influence of a time-periodic space-homogenous external field
$E(t)=E_{ac}\sin(\omega t)$ \cite{Ham}, to which we add an external
dc bias $E_{b}$:
\begin{eqnarray}
\dot{x}=p,~~\dot{p}=\sin(2\pi x)+E_{ac}\sin(\omega t)-E_{b}
\label{eq:model}
\end{eqnarray}

Due to time and space periodicity of the system we can map
the original three-dimensional phase space $(x, p, t)$ onto a
two-dimensional cylinder, $\texttt{T}^{2}=(x \; mod 1,p)$, by using  the
stroboscopic Poincar$\acute{e}$ section after each period $T=2\pi/\omega$,
 cf. Fig.1(a). Although the dynamics is periodic
in $x$ direction, it contains the complete information about
transport in the extended system (1). The velocity $\dot{x}=p$
along a trajectory is the same for both cases.

In the case of zero dc bias, $E_{b}=0$, the phase space of the
system is characterized by the presence of a stochastic layer
which originates from the destroyed separatrix of the undriven
case $E_{ac}=0$ \cite{Ham}. The chaotic layer is confined by
transporting KAM-tori, which originate from perturbed trajectories
of  particles with large kinetic energies. These tori are
\textit{noncontractible} (since they cannot be continuously
contracted to a point on $\texttt{T}^{2})$, and separate the
cylinder.

The stochastic layer is not uniform and contains regular invariant
manifolds. There exists a whole hierarchy of embedded regular
islands, which are filled by \textit{contractible} tori, cf.
Fig.1(a). The centers of the islands contain elliptic periodic
orbits $\hat{\textbf{X}}_{T_{p}}(t)=\{x_{T_{p}}(t),
p_{T_{p}}(t)\}$ with the period $T_{p}=kT, ~k=1,2,...$ and integer
shifting distance $L$, such that
\begin{equation}
x_{T_{p}}(t+T_{p})=x_{T_{p}}(t)+L,~~p_{T_{p}}(t+T_{p})=p_{T_{p}}(t)
\label{eq:PO}
\end{equation}
 The transport properties of such an orbit (and of all trajectories
inside the corresponding island) are characterized by a winding
number $\upsilon=L/T_p$. If $\upsilon \neq 0$ then that PO and the
corresponding island are transporting. In addition, within the
chaotic sea there is an infinite number of unstable periodic
orbits (UPO) with different periods and winding numbers. UPOs are
unstable with respect to small perturbations and they are not
isolated from the chaotic layer. Finally \textit{cantori}
\cite{Cantor} exist which generate semipenetrable barriers in
phase space.

Eq.(\ref{eq:model}) is invariant under time reversal
\begin{equation}
S: t \rightarrow -t+T/2, ~~x \rightarrow x, ~~p \rightarrow -p,
\label{eq:sym}
\end{equation}
which changes the sign of the current $J=v$. Thus, transporting
invariant POs and islands appear as symmetry-related pairs
in phase space with opposite winding numbers.

Contrary to the common expectation that all trajectories
acquire unbounded acceleration for nonzero dc bias $E_{b} > 0$,
recent numerical evidence
tells
that this is not true for all trajectories \cite{Ketzmerick}.
In Fig.1(b) we plot the escape time $T_{esc}$
to be accelerated below the threshold $p_t=-10$ as a function
of initial conditions $\{x_{0}, p_{0}\}$ for nonzero $E_{b}$.
The window $\{x_{0}, p_{0}\}$ coincides with Fig.1(a).
The escape time is reached by a trajectory if $p(T_{es})=-10$,
and we integrate up to $t=200$. While most of the chaotic layer in Fig.1(a) is quickly accelerated,
regular islands persist and their trajectories are in fact not accelerated at all.
Moreover we observe phase space regions with delayed acceleration.
These trajectories stick for large time to the persisting regular islands.

{Noncontractible} KAM-tori do not survive in the biased Hamiltonian
system.
If there exists at least one accelerating
trajectory,
then KAM-tori do not persist \cite{Meiss}.
Such accelerating trajectories always exist for $E_b \neq 0$,
both with $E_{ac}= 0$ and $E_{ac}\neq 0$.

\begin{figure}[t]
\includegraphics[width=0.95\linewidth,angle=0]{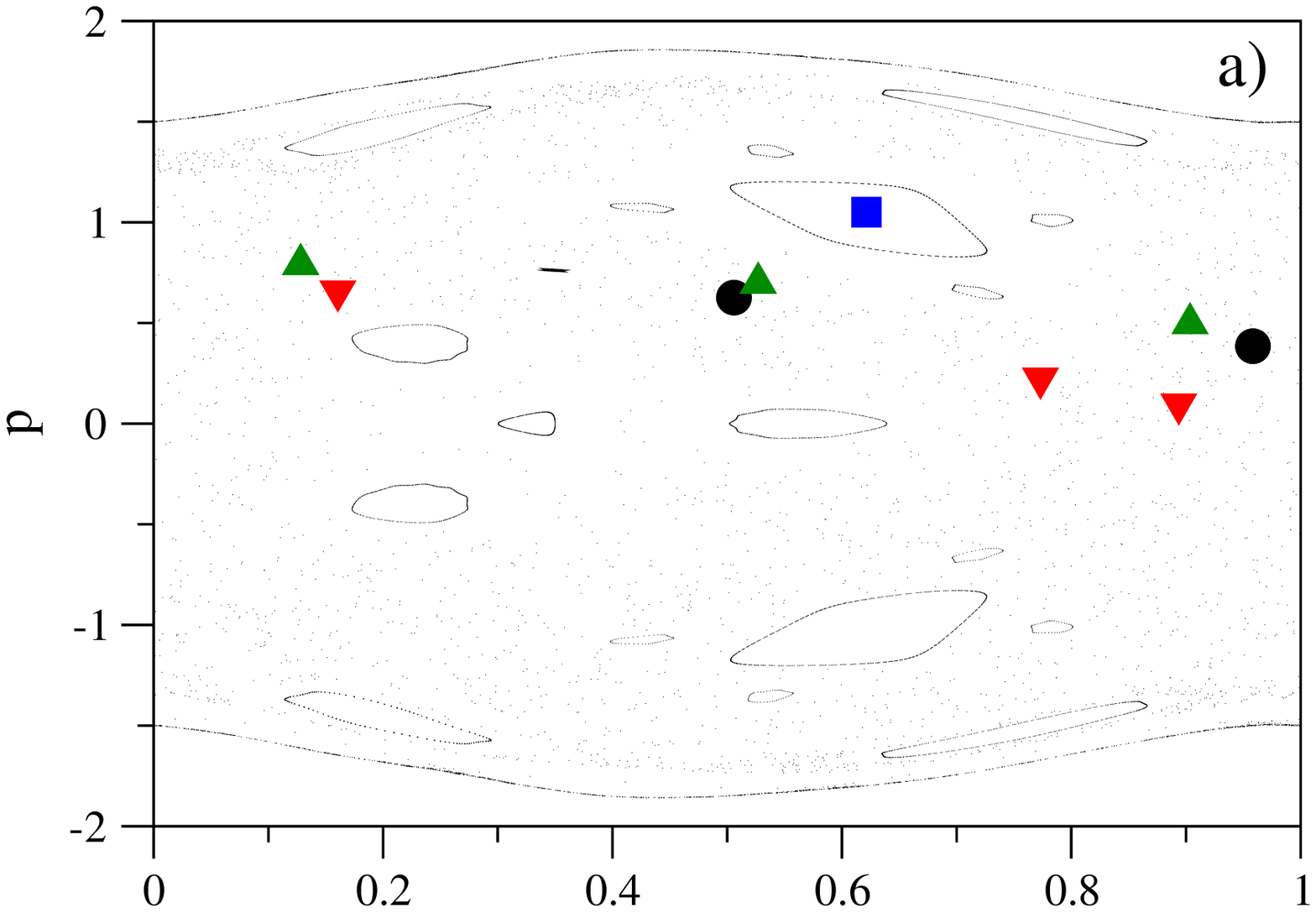}
\includegraphics[width=8cm,angle=-0]{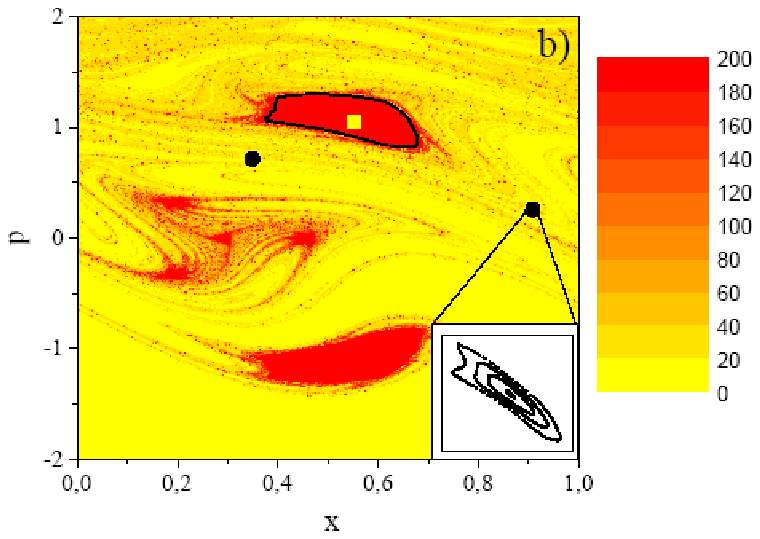}
\caption{(a) Poincar$\acute{e}$ section for the system Eq.(1), $\omega
= 2\pi, T=1, E_{ac} =5.8$, for (a) $E_{b}=0$ and (b) $E_{b}=0.183$.
Several POs are shown: one stable ($k=1,
L=1$)(squares), and several unstable POs ($k=2, L=1$)(circles),
($k=3, L=2$)(upward
triangles),  and ($k=3,
L=1$)(downward triangles). The island group near $p=0$ has zero winding number
$\upsilon=0$.
(b) Time $T_{esc}$ to be accelerated below the threshold $p_{t}=-10$
as a function of the initial conditions in phase space. Trajectories from the
islands are not accelerated at all.
Inset in Fig.1(b): Poincare section zoom of phase space
structure showing the transformation of an UPO into a stable PO
with a surrounding invariant regular island. } \label{fig:conv}
\end{figure}

Let us give analytical proof that POs at $E_b=0$ persist for
nonzero $E_b$. In order to formulate the proof we define the
stability property of a PO $\hat{\textbf{X}}_{T_{p}}(t)$. Towards
this goal we linearize the phase space flow around it, and map it
onto itself by integrating over one period $T_{p}$. The resulting
$2 \times 2$ symplectic Floquet matrix $\textbf{M}$ has
eigenvalues (Floquet multipliers) $\lambda_{1}$ and $\lambda_{2}$
with $\lambda_{1}\lambda_{2}=1$ \cite{Gut}. For a stable PO, both
multipliers are located on the unit circle, while for an UPO, both
multipliers are located either both on the negative or positive
real axis.

Denote a solution of (\ref{eq:model})
with initial conditions $\textbf{X}_{0}=\{x_{0}, p_{0}\}$ at $t=0$ by
$\hat{\textbf{X}}(t,\textbf{X}_{0},E_{b})=\{x(t,x_{0},
p_{0},E_{b}),p(t,x_{0}, p_{0},E_{b})\}$.
If $\textbf{X}_{0}$ is close enough to
$\hat{\textbf{X}}_{T_{p}}(0)$ and $E_{b}$ is sufficiently small, then
the solution $\hat{\textbf{X}}(t,X_{0},E_{b})$ on the
interval $[0,T_{p}]$ is close to the unperturbed PO.
$\hat{\textbf{X}}$ is $T_{p}$-periodic
if and only if the vector function
\begin{equation}
\textbf{F}(\textbf{X}_{0}, E_{b})= \left( \begin{array}{cccc}
x(T_{p},x_{0},p_{0},E_{b})-x_{0}-L\\
p(T_p,x_{0}, p_{0},E_{b})-p_{0}\\
\end{array} \right)=\textbf{0}.
\label{eq:Jacobian}
\end{equation}
Since $\hat{\textbf{X}}(t,x_{{T_{p}}}(0), p_{{T_{p}}}(0),0)\equiv
\hat{\textbf{X}}_{T_{p}}(t)$, Eq.~(\ref{eq:Jacobian}) is satisfied
by $\hat{\textbf{X}}_{{T_{p}}}(0)$ at $E_{b}=0$. By the Implicit
Function Theorem \cite{Schwartz}, if the Jacobian of the vector
function $\textbf{F}(\textbf{X}_{0}, E_{b})$ with respect to
$\textbf{X}_{0}$ is non-zero at $\hat{\textbf{X}}_{{T_{p}}}(0)$,
$E_{b}=0$, then the solution $\hat{\textbf{X}}_{{T_{p}}}(0)$ can
be continued to $E_{b}\neq 0$,
$\textbf{X}_{0}(E_{b})=\{x_{0}(E_{b}), p_{0}(E_{b})\}$ which
corresponds to a $T_{p}$-periodic orbit of the perturbed system.
The Jacobian of the function (\ref{eq:Jacobian}) at
$\hat{\textbf{X}}_{T_{p}}(0)$ is $ \det
\partial \textbf{F} /
\partial \textbf{X} = \det [ \textbf{M}  -\textbf{I}]$, where
$\textbf{I}$ is the unity matrix. The Jacobian is nonzero if
$\lambda_{1,2}\neq 1$ for $\textbf{M}$. Thus, almost all POs of
the unperturbed system persist for nonzero dc bias since
generically their Floquet multipliers are not equal to unity. POs
can be continued up to a value of $E_b$ where their Floquet
multipliers collide at +1.
\begin{figure}[t]
\includegraphics[width=0.8\linewidth,angle=0]{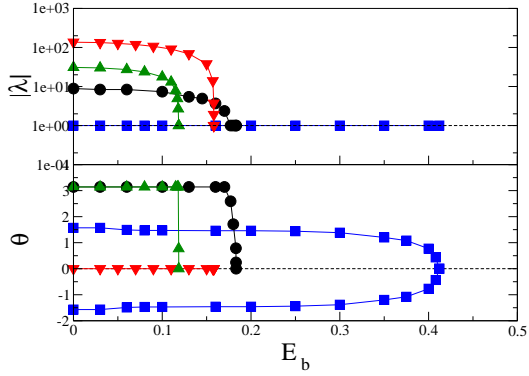}
\caption{ Variation of Floquet multipliers,
$\lambda_{j}=|\lambda_{j}|e^{i\theta_{j}}$, of PO's {\it vs} bias
$E_{b}$ for  Eq.(1) with parameters and symbols as in Fig.(1). We show both
multipliers for stable PO (squares) and
$|\lambda_{1}| > 1$ for unstable POs.} \label{alphascal8190}
\end{figure}
Linearly stable POs in Hamiltonian systems are always enclosed by
quasiperiodic tori, which form a regular island. Thus,
the transporting contractible islands also persist for nonzero  dc bias.
Note that the symmetry (\ref{eq:sym}) persists for nonzero
$E_{b}$. Consequently any invariant transporting manifolds which
persist for nonzero $E_b$ come in pairs and stationary currents
occur in both directions \cite{Yevtushenko}.

We expect that for large enough dc bias $E_{b} \gg 1$, a
transporting PO will disappear. This occurs when its Floquet
multipliers collide at $+1$,
$\lambda_{1}(E^{c}_{b})=\lambda_{2}(E^{c}_{b})=1$. This is a
saddle-center bifurcation \cite{Wiggins}, when two POs of the same
period coalesce, one of them being initially stable and the other
one unstable. Since stable POs are enclosed by regular islands,
this can happen only when the corresponding island surrounding the
stable PO shrinks to zero.

In Fig.1(a) we depict the phase space structure of
Eq.(\ref{eq:model}) for $\omega=2\pi, E_{ac}=5.8$ and $E_b=0$.
Using the Newton method, cf.  e.g. \cite{Wiggins}, we find stable
and unstable POs (various symbols in Fig.1), and continue them to nonzero values of $E_b$.
Stable POs exist up to $E_{b} \approx 0.412$.
The evolution of the corresponding Floquet multipliers
$\lambda=|\lambda|\exp(i\theta)$ is shown in Fig.2.

Transporting UPOs become more stable with increasing dc bias,
since they have to acquire $\lambda_{1,2}=1$ at a critical dc bias
value before disappearing. Moreover, UPOs with negative real
Floquet multipliers have to become stable POs before a critical
bifurcation happens, since the only way to move the multipliers
from the negative real axis to +1 is to transport them to -1
 and subsequently to shift them along
the unit circle. During that last stage an originally unstable PO
transforms into a stable PO, and a new transporting regular island
emerges in phase space. Thus, along with an overall shrinking of
the size of transporting regular islands with increasing $E_b$,
new islands are born as well. In the inset in Fig1(b) a zoom of
the Poincare section shows such a case. For $E_b=0$ we found an
UPO with $T_{p}=2$ and winding number $\upsilon=1/2$ and
$\lambda_{1}\approx-8.365, \lambda_{2}=1/\lambda_{1}\approx
-0.119$. This orbit becomes stable at $E_{b} \approx 0.178$ (when
the multipliers collide at $-1$ and enter the unit circle) and
disappears at $E_{b} \approx 0.185$, when
$\lambda_{1}=\lambda_{2}=1$. In between these two bias values the
now stable PO is surrounded by a regular island of finite size.
\begin{figure}[t]
\includegraphics[width=0.9\linewidth,angle=0]{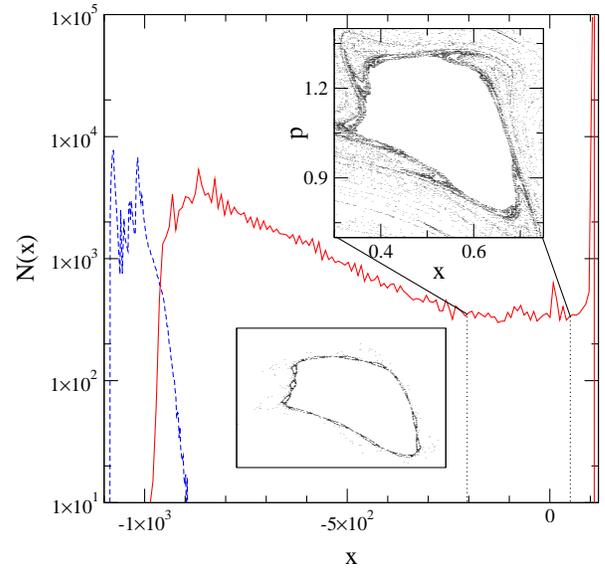}
\caption{Solid line: spatial distribution $N(x)$ after $t=100$
for
$\omega=2\pi, E_{ac}=5.8$ and $E_{b}=0.183$
and when $2.5\cdot 10^{4}$ trajectories started out from the uniform
distribution in the  rectangle $(x\in [0.3, 0.75], p\in [0.7, 1.4])$,
which enclose the transporting island with $\upsilon=1$. The upper
inset shows initial conditions which correspond to the
tail of $N(x)$, $x\in [-200,50]$. Lower inset depicts the Poincare section
of a single
trajectory that started out in the "sticky" vicinity of the island.
Dashed line: distribution $N(x)$  with initial conditions shifted by $\Delta p_{0}=-2.2$ and additional shift in $x$ (see text).}
\label{alphascal8191}
\end{figure}

For any PO the momentum  $p(t)$ is a
periodic function and  can be represented in the form $p(t)=\upsilon
+ p^{0}(t)$, such that $p^{0}(t+T_{p})=p^{0}(t)$, $\langle
p^{0}(t) \rangle_{T_{p}}=0$. Then we obtain
the energy balance equation:
\begin{equation}
\upsilon E_{b} = \langle p^{0}(t)E_{ac}(t) \rangle_{T_{p}}.
\label{eq:balance}
\end{equation}
An external ac-field mimics
an additional effective friction in the nonequilibrium
Nos$\acute{e}$-Hoover oscillator \cite{Nose}: it pumps  energy
into the system in order to compensate the work against bias (for POs
with positive $\upsilon$) and acts as an energy sink to compensate
the acceleration in the case of current along a bias (for POs with
negative $\upsilon$). But in our case, at variance with the
Nos$\acute{e}$-Hoover system, the generating equations
are Hamiltonian and dissipationless.

All trajectories except the regular islands and the POs will be
accelerated without bound for nonzero $E_b$. Nevertheless, the
mixed phase space structure for $E_b=0$ leads to a strongly
nonuniform acceleration in time for trajectories with different
initial conditions. That can be already observed in Fig.1(b).
Indeed, for $E_b=0$ chaotic trajectories in the vicinity of
regular islands may stick to the island boundary for long times
and produce a unidirectional flight with the velocity $\upsilon$
due to the \textit{stickiness} effect \cite{Klafter}. The origin of
such a behavior is the presence of partial barriers in phase space
formed by cantori \cite{Cantor}. This feature is also
observed on finite times for nonzero $E_b$. Trajectories may stick
for long times to regular island boundaries without acceleration
(see Poincar$\acute{e}$ section in lower inset in Fig.3). Only
after some rather large escape times $T_{esc}$ such a trajectory
will eventually leave the island vicinity and suddenly accelerate. In Fig.3
(solid line) we compute the distribution $N(x)$ of displacements
of trajectories with initial conditions uniformly covering a phase
space part which includes a transporting island. We observe that
the
transient time $T_{esc}$ can be extremely long for initial
conditions in the  island's vicinities (see also
Fig.1(b)). In Fig.3 the right peak corresponds to initial
conditions inside the island (no acceleration for all times). The
left front peak corresponds to initial conditions that are
accelerated right from the beginning.
In the absence of trajectories which stick to regular islands
with a broad distribution of escape times, the two peaks would
be separated by a depleted region, i.e. a gap.
Instead we observe a smeared out distribution which
corresponds to trajectories with long transient sticking to the
island boundary. Indeed, initial conditions which correspond to the
gap region stick to the island (upper inset in Fig.3).
A shift of the initial conditions by $\Delta p_{0}=-2.2$ down into
the homogenous part of the phase space in Fig.1(b) yields a
distribution $N(x)$ which contains a single peak due to immediate
acceleration (Fig.3, dashed line). Note that we shifted $N(x)$ here by $\Delta x_{0}=220$
which occurs for particles with $V(x)=0$ and $E_{ac}=0$ if the
initial momentum is shifted by $\Delta p_{0}=-2.2$.

In conclusion, we show that stationary, dissipationless transport
is sustained by a nonzero dc bias. The dc bias provides a unique
possibility to separate the different parts of a mixed phase
space, since it accelerates the chaotic part away and leaves the
regular invariant manifolds basically untouched. Moreover, even
within the accelerated part, proper distribution functions after
finite acceleration times reveal the intriguing properties of
surviving cantori and sticking. We give an analytical proof for the
persistence of periodic orbits and thus for regular islands. We
found evidence for the persistence of cantori as well. Our results
are not limited to the case in Eq.(\ref{eq:model}) and are valid as well
for the general case of a potential $U(t,x)=U(t+T,x)=U(t,x+L)$.
The presence of weak dissipation will not change the situation
drastically. Since it can be treated as a smooth perturbation, POs
will persist for nonzero damping. Stable POs become attractors
(limit cycles), and unstable POs saddles \cite{Gut}. A side result
 is then that an infinite number of stable POs for zero
damping will lead to an infinite number of attractors for
infinitesimally weak damping \cite{Feudel}.

A possible experimental realization
can be obtained with
cold atoms in a standing wave potential
\cite{cold}, where an additional dc-bias can be realized by a
gravitational force. In such a case, all of the above
results could be verified. The persistence of transporting regular
islands and POs can be used for the preparation of monochromatic
matter waves.

\end{document}